\documentclass{ws-ijmpcs}

\begin{document}

%
\catchline{}{}{}{}{}
%

\title{Heavy Quark Symmetries: Molecular Partners of the $X(3872)$ and $Z_{b}(10610)/Z_{b}'(10650)$.}

\author{Feng-Kun Guo}

\address{Helmholtz-Institut f\"ur Strahlen- und
             Kernphysik and Bethe Center for Theoretical Physics, \\
             Universit\"at Bonn,  D-53115 Bonn, Germany}

\author{Carlos Hidalgo-Duque, Juan Nieves}

\address{Instituto de F\'isica Corpuscular (IFIC),
             Centro Mixto CSIC-Universidad de Valencia,
             Institutos de Investigaci\'on de Paterna,
             Aptd. 22085, E-46071 Valencia, Spain}



\author{Manuel Pav\'on Valderrama}

\address{Institut de Physique Nucl\'eaire,
             Universit\'e Paris-Sud, IN2P3/CNRS,
             F-91406 Orsay Cedex, France}

\maketitle

\begin{history}
\received{Day Month Year}
\revised{Day Month Year}
\end{history}

\begin{abstract}

In this work, we have used an Effective Field Theory (EFT) framework based on Heavy Quark Spin (HQSS), Heavy Flavour (HFS) and Heavy Antiquark-Diquark symmetries (HADS). Using a standard lagrangian for the heavy meson-heavy antimeson system, we fit the counter-terms of the model to predict some promising experimental data that can be interpreted as heavy meson-heavy antimeson molecules, that is, the $X(3872)$ and the $Z_{b}(10610)/Z'_{b}(10650)$. Next, and, taking advantage of HADS, we use the same lagrangian to explore the consequences for heavy meson-doubly heavy baryon molecules, which can also be interpreted as triply heavy pentaquarks.

\end{abstract}

\ccode{PACS numbers: 03.65.Ge,13.75.Lb,14.40Pq,14.40Rt}

\section{Introduction}	

Since the mid 1970's the existence of heavy hadronic molecules (composed by a pair of heavy hadrons instead of a pair of heavy quarks) has been theorized\cite{Voloshin:1976ap}. This assumption was made based on the similarities between the heavy meson-heavy antimeson system and the deuteron. However, it was not until the discovery of the $X(3872)$\cite{Choi:2003ue} by the Belle Collaboration, in 2003, that the first experimental data that could fit into that molecular scheme was found. Since then, many other XYZ states have been found, being the $Z_{b}(10610)/Z'_{b}(10650)$ also natural candidates to have a molecular structure..

Besides, the $m_{Q} \rightarrow \infty$ limit of QCD simplifies the theory so a set of symmetries are induced. Probably, the most important symmetries induced in this limit are HQSS, HFS, and HADS. We make use of them, along with the assumptions of the $X(3872)$ and the $Z_{b}(10610)/Z'_{b}(10650)$ to be heavy hadronic molecules, to obtain a family of heavy meson-doubly heavy baryons that could also be interpreted as triply heavy pentaquarks.

This proceeding is organized as follows. First we briefly introduced our EFT based on HQSS and HFS that we will use in the analysis of the $X(3872)$ and the $Z_{b}(10610)/Z'_{b}(10650)$. Second, we will discuss HADS and its implications. Finally, our results will be shown in Table \ref{tab:predictions}.

\section{Effective Field Theory for Heavy Mesonic Molecules and Lippmann-Schwinger Equation.}

In this work we are following the scheme described in \cite{HidalgoDuque:2012pq} where all sort of details can be found. At lowest order, HQSS and HFS impose that the dynamics of the model does not depend on either the mass or the spin of the heavy quark. Taking this into account, the most general potential that describes the dynamics of the heavy meson-antimeson pair depends only in four Low Energy Constants or counter-terms (LECs), up to corrections of the order $\mathcal{O}\left( \frac{1}{m_{Q}}\right)$:
\begin{eqnarray}
\nonumber{}
V_{4} & = &+  \frac{C_{A}}{4}~ Tr\left[\bar{H}^{(Q)}{H}^{(Q)} \gamma_{\mu} \right] Tr\left[{H}^{(\bar{Q})} \bar{H}^{(\bar{Q})} \gamma^{\mu} \right] + \\
\nonumber{}
&& + \frac{C_{A}^{\lambda}}{4}~ Tr\left[\bar{H}^{(Q)}_{a} \lambda^{i}_{ab} {H}^{(Q)}_{b} \gamma_{\mu} \right] Tr\left[{H}^{(\bar{Q})}_{c} \lambda^{i}_{cd}\bar{H}^{(\bar{Q})}_{d} \gamma^{\mu} \right]  + \\
\nonumber{}
&& + \frac{ C_{B}}{4}~ Tr\left[\bar{H}^{(Q)}{H}^{(Q)} \gamma_{\mu} \gamma_{5} \right] Tr\left[{H}^{(\bar{Q})} \bar{H}^{(\bar{Q})} \gamma^{\mu} \gamma_{5}\right] + \\
&& +  \frac{C_{B}^{\lambda}}{4}~ Tr\left[\bar{H}^{(Q)}_{a} \lambda_{ab}^{j}{H}^{(Q)}_{b} \gamma_{\mu} \gamma_{5} \right] Tr\left[{H}^{(\bar{Q})}_{c} \lambda^{j}_{cd} \bar{H}^{(\bar{Q})}_{d} \gamma^{\mu} \gamma_{5}\right]
\end{eqnarray}
being $\lambda$ the Gell-Mann matrices and $H^{Q(\bar{Q})}$ the meson (antimeson) field in the charm sector (and viceversa in the bottom sector). Moreover, the four LECs will be rewritten through a linear combination into $C_{0A}$, $C_{0B}$, $C_{1A}$ and $C_{1B}$ for notation.

\begin{table}[]
\tbl{\label{tab:potentials} LO potentials and quantum numbers for various
doubly-heavy baryon--heavy meson systems.
}
{\begin{tabular}{l || c c c c c }
       States & $\Xi_{Q_1Q_2}P$ & $\Xi_{Q_1Q_2}P^*$ &
       $\Xi_{Q_1Q_2}P^*$  & $\Xi_{Q_1Q_2 }^*P$
       & $\Xi_{Q_1Q_2}^*P^*$  \\ \hline\hline
       $J^{P}$ & $\frac12^-$ & $\frac12^-$ & $\frac32^-$  & $\frac32^-$ &
       $\frac12^-$  \\  \vspace{0.3cm}
       $V^{\rm LO}$ & $C_{Ia}$ & $C_{Ia}+\frac23 C_{Ib}$ &
       $C_{Ia}-\frac13 C_{Ib}$ & $C_{Ia}$ & $C_{Ia}-\frac53 C_{Ib}$ \\

	States & $\Xi_{Q_1Q_2}^*P^*$ & $\Xi_{Q_1Q_2}^*P^*$ & $\Xi_{bc}'P$ &
       $\Xi_{bc}'P^*$ &       $\Xi_{bc}'P^*$ \\ \hline\hline
$J^{P}$ & $\frac32^-$ & $\frac52^-$ & $\frac12^-$ & $\frac12^-$ &
       $\frac32^-$ \\
	$V^{\rm LO}$ &  $C_{Ia}-\frac23 C_{Ib}$ & $C_{Ia}+C_{Ib}$ & $C_{Ia}$ & $C_{Ia}-2C_{Ib}$ &
       $C_{Ia}+C_{Ib}$ 
   \end{tabular} }
\end{table}

\begin{table}[]
\tbl{\label{tab:predictions}
Predictions of the doubly-heavy baryon--heavy meson
molecules.
Results are given in terms of $M_{\rm th}$ for simplicity, and all masses are given in units
of MeV.  When we decrease the
strength of the potential to account for the various uncertainties, in some 
cases (marked with $\dagger$ in the table)  the bound state pole reaches the 
threshold and the state becomes virtual. The cases with a virtual state pole 
at the central value are marked by [V], for which $\dagger\dagger$ 
means that the pole 
evolves into a bound state one and N/A means that the pole is far from the 
threshold with a momentum larger than 1~GeV so that it is both undetectable and 
beyond the EFT range.}
{    \begin{tabular}{l c c c c c}
       State & $I(J^{P})$ & $V^{\rm LO}$ & Thresholds &
       Mass ($\Lambda=0.5$ GeV) & Mass ($\Lambda=1$ GeV) \\ \hline 
       $\Xi_{cc}^* D^*$ & $0({\frac{5}{2}}^-)$ & $C_{0a}+C_{0b}$ & $5715$
       & $\left( M_{\rm th} - 10\right)^{+10}_{-15}$ & $\left(M_{\rm th} - 
19\right)^{\dagger}_{-44}$ \\
       $\Xi_{cc}^* \bar{B}^*$ & $0({\frac{5}{2}}^-)$ & $C_{0a}+C_{0b}$ &
       $9031$ & $\left(M_{\rm th} - 21\right)^{+16}_{-19}$ & $\left(M_{\rm th} - 
53\right)^{+45}_{-59}$ \\
       $\Xi_{bb}^* D^*$ & $0({\frac{5}{2}}^-)$ & $C_{0a}+C_{0b}$ & $12160$
       & $\left(M_{\rm th} - 15\right)^{+9}_{-11}$ & $\left(M_{\rm th} - 
35\right)^{+25}_{-31}$ \\
       $\Xi_{bb}^* \bar{B}^*$ & $0({\frac{5}{2}}^-)$
       & $C_{0a}+C_{0b}$ & $15476$
       & $\left(M_{\rm th} - 29\right)^{+12}_{-13}$ & $\left(M_{\rm th} - 
83\right)^{+38}_{-40}$ \\
       $\Xi_{bc}' D^*$ & $0({\frac{3}{2}}^-)$ & $C_{0a}+C_{0b}$ & $8967$
       & $\left(M_{\rm th} - 14\right)^{+11}_{-13}$ & $\left(M_{\rm th} - 
30\right)^{+27}_{-40}$ \\
       $\Xi_{bc}' \bar{B}^*$ & $0({\frac{3}{2}}^-)$
       & $C_{0a}+C_{0b}$ & $12283$
       & $\left(M_{\rm th} - 27\right)^{+15}_{-16}$ & $\left(M_{\rm th} - 
74\right)^{+45}_{-51}$ \\
$\Xi_{bc}^* D^*$ & $0({\frac{5}{2}}^-)$ & $C_{0a}+C_{0b}$ & $9005$
& $\left(M_{\rm th} - 14\right)^{+11}_{-13}$ & $\left(M_{\rm th} -
30\right)^{+27}_{-40}$ \\ 
$\Xi_{bc}^* \bar{B}^*$ & $0({\frac{5}{2}}^-)$ & $C_{0a}+C_{0b}$ & $12321$
& $\left(M_{\rm th} -27\right)^{+15}_{-16}$ & $\left(M_{\rm th} 
-74\right)^{+46}_{-51}$ \\    \hline 
       $\Xi_{bb}\bar{B}$ & $1({\frac{1}{2}}^-)$ & $C_{1a}$ & $15406$
       & $\left(M_{\rm th} - 0.3\right)_{-2.5}^{\dagger}$ &
       $\left(M_{\rm th} -  12\right)^{+11}_{-15}$ \\
       $\Xi_{bb}\bar{B}^*$ & $1({\frac{1}{2}}^-)$ &
       $C_{1a}+\frac{2}{3}\,C_{1b}$ & $15452$
       & ($M_{\rm th}-0.9$)[V]$_{\dagger\dagger}^\text{N/A}$
       & $\left(M_{\rm th} - 16\right)^{+14}_{-17}$ \\
       $\Xi_{bb}\bar{B}^*$ & $1({\frac{3}{2}}^-)$
       & $C_{1a}-\frac{1}{3}\,C_{1b}$ & $15452$
       & $\left(M_{\rm th} - 1.2\right)^{\dagger}_{-2.9}$
       & $\left(M_{\rm th} - 10\right)^{+9}_{-13}$ \\
       $\Xi_{bb}^*\bar{B}$ & $1({\frac{3}{2}}^-)$ & $C_{1a}$ & $15430$
       & $\left(M_{\rm th} - 0.3\right)^{\dagger}_{-2.4}$
       & $\left(M_{\rm th} - 12\right)^{+11}_{-13}$ \\
       $\Xi_{bb}^*\bar{B}^*$ & $1({\frac{1}{2}}^-)$
       & $C_{1a}-\frac{5}{3}\,C_{1b}$ & $15476$
       & $\left(M_{\rm th} - 8\right)^{+8}_{-7}$
       & $\left(M_{\rm th} - 5\right)^{\dagger}_{-8}$ \\
       $\Xi_{bb}^*\bar{B}^*$ & $1({\frac{3}{2}}^-)$ & 
$C_{1a}-\frac{2}{3}\,C_{1b}$ & $15476$
       & $\left(M_{\rm th} - 2.5\right)^{\dagger}_{-3.6}$
       & $\left(M_{\rm th} -9\right)^{+9}_{-11}$ \\
       $\Xi_{bb}^*\bar{B}^*$ & $1({\frac{5}{2}}^-)$ & $C_{1a}+C_{1b}$ & $15476$
       & $\left(M_{\rm th}-4.3\right)$[V]$_{+3.3}^\text{N/A}$
       & $\left(M_{\rm th} - 18\right)^{+17}_{-19}$ \\
   \end{tabular} }
\end{table}

These four LECs will be fitted to reproduce some experimental data in our scheme. In this framework, bound states will be found by solving the Lippmann-Schwinger equation as they will appear as poles in the T-matrix: $T = V + V ~G~T$, and the ultraviolet divergences of the loop function are treated introducing a Gaussian regulator in the propagator and in the potential such as:

\begin{equation}
\left<\vec{p}|V|\vec{p}~'\right> = v ~e^{-\vec{p}^{2}/\Lambda^{2}}~e^{-\vec{p}'^{2}/\Lambda^{2}} ~~~\Rightarrow ~~~ G = \int \frac{d^{3}\vec{k}}{(2\pi)^{3}}  \frac{ e^{-2\vec{k}^{2}/\Lambda^{2}} } {E - m_{1}-m_{2} - \frac{k^{2}}{2\mu} }
\end{equation}

These assumptions determine three linear combinations of the LECs, that is: $C_{0X} = C_{0A} + C_{0B}$ and $C_{1X} = C_{1A} + C_{1B}$ which in turn are determined by the $X(3872)$ (for more details, see \cite{HidalgoDuque:2012pq}) and $C_{1Z} = C_{1A} - C_{1B}$ by the $Z_{b}(10610)/Z_{b}'(10650)$ resonances [\cite{Guo:2013sya,Guo:2013xga}].

\section{Heavy Antiquark-Diquark Symmetry and Results.}

Up to now, we have only established an EFT that analyzes heavy meson-heavy antimeson molecules. In order to use this approach to different systems we will take advantage of the Heavy Antiquark-Diquark Symmetry (HADS). This $m_{Q}\rightarrow \infty$ limit symmetry, first introduced by Savage and Wise, states that a heavy diquark behaves as a heavy antiquark up to corrections of the order $\mathcal{O}\left(\frac{1}{m_{Q}v}\right)$, , being v the velocity of the heavy quarks.

Furthermore, since the dynamics of our EFT only depends on the light degrees of freedom, that are the same than in the heavy meson-heavy antimeson system, we can make use of some Racah algebra (similar to \cite{Voloshin:2011qa}) to obtain the potentials in every possible channels, which are displayed in Table \ref{tab:potentials}. The $\Xi'$ states are those where heavy quarks in the baryon are coupled to $S_{Q_{baryon}} = 0$ (which is forbidden if the two quarks are the same because of the Pauli's principle of exclusion).

Then we just have to solve the Lippmann-Schwinger equation in each channel using the LECs we have previously fitted to obtain the results of Table \ref{tab:predictions}. The isoscalar states are related to the $X(3872)$. The isovector states are determined by the $Z_b(10610,10650)$ and the isovector component of the $X(3872)$.

The sources of error in the analysis are: the masses of the $X(3872)$ and $Z_{b}$ resonances, the ratio of the $X(3872)$ amplitude decays (calculated in \cite{Hanhart:2011tn}) and the two EFT expansions used in this work. The errors for HQSS are taken to be $20\%~(7\%)$ in the charm (bottom) sector and $40\%~(30\%)~[20\%]$ in the charm (charm-bottom) [bottom] sector for HADS. Then, an unique error is obtained by adding in quadratures all different sources.

\section{Conclusions}

As a summary, we can conclude that our analysis based on several QCD symmetries in the $m_{Q}\rightarrow \infty$ limit predicts the existence of several heavy meson-doubly heavy baryon molecular partners of the $X(3872)$ and the $Z_{b}(10610)/Z'_{b}(10650)$.

This same effective field theory approach could also be extended to study doubly heavy baryon-double heavy antibaryon molecular systems in the future. 

\section*{Acknowledgments}

F.-K.G. acknowledges the Theory Division of IHEP in Beijing, where part of the work was done, for the hospitality. C. H.-D. thanks the JAE-CSIC Program. This work is supported in part by
the DFG and the NSFC through funds provided to the Sino-German CRC 110
``Symmetries and the Emergence of Structure in QCD'', by the NSFC
(Grant No. 11165005), by the Spanish Ministerio de Econom\'\i a y
Competitividad and European FEDER funds under the contract
FIS2011-28853-C02-02 and the Spanish Consolider-Ingenio 2010 Programme
CPAN (CSD2007-00042), by Generalitat Valenciana under contract
PROMETEO/2009/0090 and by the EU HadronPhysics2 project, grant
agreement no. 227431.

\end{document}